\documentclass[%
reprint,
 amsmath,amssymb,
 aps,
]{revtex4-2}

\usepackage{graphicx}
\usepackage{dcolumn}
\usepackage{bm}
\usepackage{physics}
\usepackage{verbatim}

\begin{document}

\preprint{APS/123-QED}

\title{Single-particle steering and nonlocality:
The consecutive Stern-Gerlach Experiments}

\author{E{ Ben\'itez Rodr\'iguez}}
\author{E {Piceno Mart\'inez}}
\author{L M {Ar\'evalo Aguilar}}\email{larevalo@fcfm.buap.mx}\affiliation{Facultad de Ciencias F\'isico Matem\'aticas, Benem\'erita Universidad Aut\'onoma de Puebla.}%

\date{\today}

\begin{abstract}
Quantum nonlocality and quantum steering are fundamental correlations of quantum systems which can not be created using classical resources only. Nonlocality describes the ability to influence the possible results of measurements carried out in  distant systems, in quantum steering Alice remotely steers Bob's state. Research in nonlocality and steering possess a fundamental interest for the development of quantum information and in many applications requiring nonlocal resources like quantum key distribution. On the other hand, the Stern-Gerlach experiment holds an important place in the history, development and teaching of quantum mechanics and quantum information.  In particular, the thought experiment of consecutive Stern-Gerlach Experiments is commonly used to exemplify the concept of non-commutativity between quantum operators. However, to the best of our knowledge, the consecutive Stern-Gerlach Experiments have not been treated in a fully quantum manner yet, and it is a widely accepted idea that atoms crossing consecutive Stern-Gerlach Experiments follow classical paths.  Here we demonstrate that two consecutive Stern-Gerlach Experiment generate nonlocality and steering, these  nonlocal effects strongly modify our usual understanding of this experiment.  Also, we discuss the implications of this result and its relation with the entanglement. This suggests the use of quantum correlations, of particles possessing mass, to generate nonlocal taks using this venerable experiment.
\end{abstract}

\maketitle


\section{Introduction}
\label{sec:Introduction}

Nonlocality, one of the fundamental features of quantum mechanics, refers to the fact that for entangled states the result of a measurement on one observable depends on the choice of measurement on the other observable --this result manifests spacelike separation-- \cite{brunner2014,wolfe20,chaves17,araujo20}; steering refers to the ability to steer a state of one subsystem by measurements made in another subsystem \cite{wiseman2007,jones07,cavalcanti09,uola20,paris20}, it is a nonlocal property that differs from nonlocality and nonseparability \cite{wiseman2007,jones07}.
Additionally, it was demonstrated that nonlocality rests in the uncertainty principle and in the steering of physical states\cite{wehner10,ramanathan18}. Besides, the nonlocality of the collapse of the wavefunction of a single particle was experimentally proved for photons  in reference \cite{fuwa15}, for atoms see reference \cite{george13} for an experimental proof of nonclassical collapses by using boxes games \cite{vaidman91}. It is worthy of mention that steering was experimentally proven in single photon experiment by Guerreiro et al. \cite{brunner16}. Additionally, it was shown that there is a strong relation between steering and joint measurability \cite{brunner14,uloa15}. Nowadays it is understood that quantum nonlocality is a fundamental resource for quantum information tasks which can not be generated by using random data only \cite{BancalDef}. Nevertheless, see the interesting discussion over quantum nonlocality given by Khrennikov \cite{khrennikov20aa,khrennikov20a,khrennikov20b}.

Moreover, although in some works it has been conjectured that single-particle entanglement does not possess nonlocal correlations \cite{karimi20,azzini20}, the fascinating fact is that these nonlocal effects could be generated using the Stern-Gerlach Experiment (SGE) by measuring the internal degree of freedom of the particle that traverses it;  i.e. by measuring $\hat{\sigma}_z$ or $\hat{\sigma}_x$, the position of the particle could manifest itself at either of two different places at (possible) spacelike distance \cite{are20}. These nonlocal properties of the SGE were associated with the spreading of the wavefunction and named single-particle steering \cite{are20}. Hence, depending on the kind of states, the spreading of the wavefunction plays also a crucial role in the nonlocality of quantum mechanics.

Furthermore, single-particle entanglement \cite{azzini20} has many applications in quantum information like quantum key distribution \cite{adhikari15,wei19}, bidirectional quantum teleportation \cite{heo15}, swapping states \cite{adhikari10} and entanglement concentration \cite{cheng11}. Recently the single-particle entanglement between the spin and orbital angular momentum (OAM) of photons was reported in metamaterials \cite{stav18} and the entanglement between polarization and OAM was studied in reference \cite{bhatti15}. Besides, as was mentioned earlier, steering was experimentally proven using the single-photon entanglement \cite{brunner16}, this effect was called single-photon steering by Brunner \footnote{Private communication} .

On the other hand, the SGE stands as a pillar for the historical development of quantum mechanics and nowadays constitutes an active field of research, regarding both its experimental capabilities and theoretical studies \cite{Ern,corrige,Lenanuevo,paris13,Barney_2019,courtnet19,roston,qureshi12,Schmidt,Robert, Machluf,Boustimi,scully,Utz,home22,Platt,Hsu,potel06,patil,Venugopalan,Hannout,Inoue2019}, and it also constitutes an important tool in the teaching of both quantum mechanics and quantum information.

The quantum phenomena that is usually introduced in quantum mechanics courses using the Stern-Gerlach experiment include the concept of spin and the non-commutativity of quantum operators \cite{SakuraiN,Townsend,LeBellac,feynman}. For the latter is of most importance the thought experiment of consecutive Stern-Gerlach Experiments (CSGE), formally introduced by Feynman in his famous lectures \cite{feynman}, regardeless of its first conception by Heisenberg in 1927 \cite{heicon}. In quantum information and computation theory the CSGE is used to exemplify the structure of the qubit and the collapse of the wavefunction \cite{wilde17,nielsen10}.

Despite its importance, there has been a lack of quantum analyses of the CSGE, at least that we are aware of. This is surprising given the amount of research dedicated to describing and studying the usual SGE; in concrete, in recent years there have been important advances describing the SGE in a completely quantum manner \cite{Ern,corrige,roston,qureshi12, paris13,Barney_2019,courtnet19,home22,Platt,Hsu,patil}. In particular, it was shown that the SGE behaves as an entangled device and not as a measurement device as it was regarded for decades \cite{Ern,corrige}. Additionally, the violation of Bell inequalities in the SGE was shown in reference \cite{epic1}, this violation of the Bell inequalities was associated with quantum nonlocality, instead of quantum contextuality, the difference is explained in reference \cite{are20}.

In this article, we carry out the task of analyzing the CSGE using the tools of quantum mechanics, and to that end we calculate the complete quantum evolution of a spin-$1/2$ particle in such consecutive configuration by means of the evolution operator method \cite{are3,are4}. This results also could serve to define a basis of comparison for a possible experimental realization of this experiment. The quantum treatment of the CSGE gives us the opportunity to study the quantum characteristics of the system, such as the quantum correlations and entanglement \cite{epic1}.\par

The paper is organized as follows. In Sec. \ref{sec:CSGEcon}, starting from the time dependent Schr\"odinger equation for the particle, we found its time evolution applying a factorized evolution operator to the initial state of the particle, similarly as done in \cite{Ern,corrige}. We calculate in this way the evolution in each of the Stern-Gerlach apparatuses that are arranged consecutively, and arrive at the result for the total evolution applying the sequential evolutions. In Sec. \ref{sec:Steering} we analyze the steering produced by changing the measurement basis.
In Sec. \ref{sec:nonloc} we describe the quantum correlations of the CSGE and we test three different Bell-type inequalities with which we found that this system is non-local. In Sec. \ref{sec:entang} we visualize the relation between the quantum correlations and the entanglement in the CSGE by means of the works of Piceno et. al. \cite{Lenanuevo} and  Roston et. al. \cite{roston}. Finally the main results obtained are recapitulated in Sec.\ref{sec:Conclusions}, that closes the paper with some concluding remarks.
\section{Dynamic evolution of the CSGE}
\label{sec:CSGEcon}
\begin{figure}[h!]
  \centering
  \includegraphics[width=86mm]{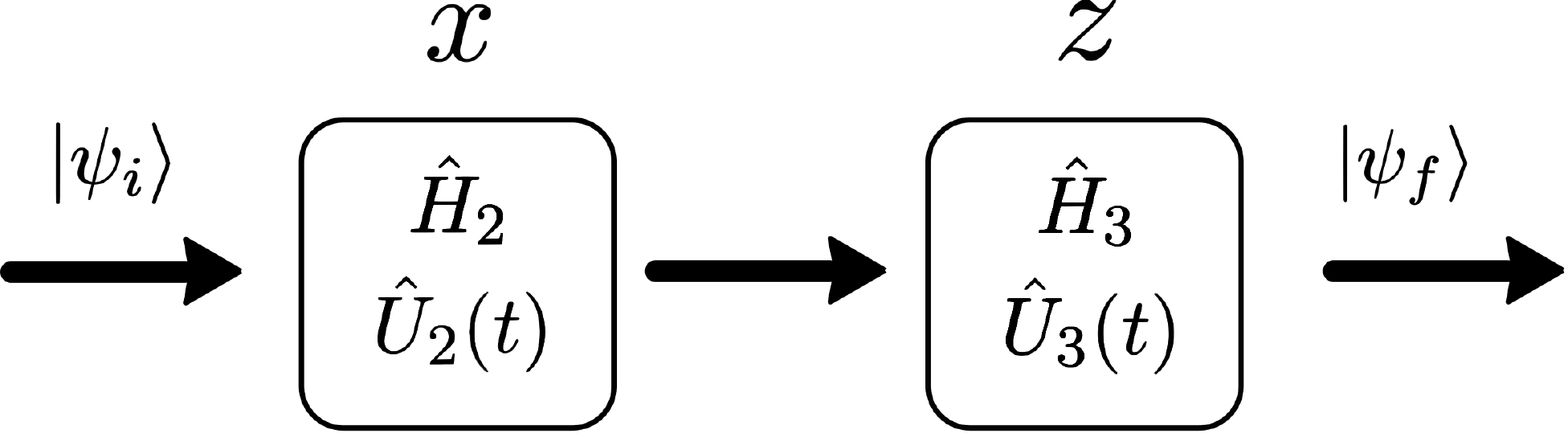}
  \caption{Scheme of the consecutive Stern-Gerlach experiments. The inhomogeneity of the magnetic field in the first experiment is in direction $x$, which is associated with a Hamiltonian, $\hat{H_{2}}$, and an evolution operator, $\hat{U_{2}}$; the inhomogeneity of the magnetic field of the second is in direction $z$, with its respective Hamiltonian, $\hat{H_{3}}$, and evolution operator $\hat{U_{3}}$.}
  \label{modif}
\end{figure}

The array of  consecutive experiments that we take into account appears schematized in FIG.\ref{modif}, this configuration has two Stern-Gerlach apparatuses -- each apparatus acts as an entangling device. The initial state entering to the SGE in $x$ direction is a state previously prepared by other SGE; that is, the initial state is given by the product of a spin up state (internal degree of freedom) with a wavepacket (external degree of freedom), which we can write as follows:
\begin{equation}\label{preparado}
\ket{\psi_{i}}=\frac{1}{(2\pi\sigma_{0}^{2})^{\frac{3}{4}}}\exp\left(-\frac{(x^{2}+y^{2}+z^{2})}{4\sigma_{0}^{2}}+ik_{y}y\right)\ket{\uparrow_z},
\end{equation}
with $\sigma_{0}$ the initial width of the wavepacket and $k_{y}$ the component $y$ of the wave vector.\par
The quantum evolution of the Stern-Gerlach experiment has already been described in \cite{Ern}, therefore, from this we know that if the initial state is a product between a gaussian wavepacket and a spin DoF the resulting evolved wavepacket widens with time, and gets translated in $z$ depending on the spin component thanks to the interaction with the inhomogeneous magnetic field when we select the spin component. With these considerations, the general packet of Eq. (\ref{preparado}) can be seen as that resulting from such preparation with a final width $\sigma_{0}$ and taking the $z$ position of such state as our origin \cite{Ern,corrige}. \par
To obtain the evolution as the state goes through the first experiment we have the associated evolution operator from \cite{Ern,corrige},
\begin{equation}\label{u2r}
\begin{split}
\hat{U}_2(t)& =\exp\left(-\frac{1}{6}\kappa\right)
	\exp\left[-\frac{i t}{2m\hbar}\left(p_y^2+p_z^2\right)\right]\\
	& \times\exp\left[-\frac{i t\mu_c}{\hbar}\left(B_2+b_2x\right)\sigma_x\right]
\exp\left(\frac{i t^2\mu_cb_2}{2m\hbar}p_x\sigma_x\right)\\
& \times\exp\left(-\frac{i t}{2m\hbar}p_x^2\right),
\end{split}
\end{equation}
with $\kappa=(i t^2\mu_{c}^{2}b_{2}^{2})/(m\hbar)$, $B_{2}$ the homogeneity parameter of the magnetic field, $b_{2}$ the inhomogeneity parameter, $\sigma_{x}$ the Pauli operator, $\mu_{c}=g{e}\hbar/(4m)$, $g$ the gyromagnetic ratio, $m$ the mass of the particle and $e$ the unit charge.\par
Applying  $\hat{U}_2(t)$ during certain time $t_{2}$ to $\ket{\psi_i}$ we have obtained the middle state $\ket{\psi_{m}(t_{2})}$. See the Appendix A for this calculation.
Next, following our description, we apply the evolution operator associated with the last SGE for a fixed time $t_{3}$, that is, the operator in $z$ direction to the state $\ket{\psi_{m}(t_{2})}$. This operator is given by
\begin{equation}\label{u3r}
\begin{split}
\hat{U}_3(t)&=\exp\left(-\frac{1}{6}\kappa\right)
	\exp\left[-\frac{i t}{2m\hbar}\left(p_x^2+p_y^2\right)\right]\\
& \times\exp\left[-\frac{i t\mu_c}{\hbar}\left(B_3+b_{3}z\right)\sigma_z\right]
	\exp\left(\frac{i t^2\mu_cb}{2m\hbar}p_z\sigma_z\right)\\
& \times \exp\left(-\frac{i t}{2m\hbar}p_z^2\right)
\end{split}
\end{equation}
Then we obtain the final state as
\begin{equation}\label{3mod3}
\ket{\psi_{f}}=\ket{\psi_{+}}\ket{\uparrow_{z}}+\ket{\psi_{-}}\ket{\downarrow_{z}},
\end{equation}
in which we define
\begin{widetext}
\begin{equation}\label{psimasmenosA}
\begin{split}
\ket{\psi_{\pm}}& =M\exp\left(\mp i\sqrt{2}k_{3}^{3}\tau_{3}z_{0}\right)\exp\left\{-\frac{\left(Z\pm\left\{\sqrt{2}k_{3}^{3}\tau_{3}^{2}+i2\sqrt{2}k_{3}^{3}\tau_{3}\left[1+i\left(\tau_{2}+\tau_{3}\right)\right]\right\}\right)^{2}}{4\left[1+i\left(\tau_{2}+\tau_{3}\right)\right]}\right\}\\
& \times\left(\exp\left(-i\sqrt{2} k_{2}^{3}\tau_{2}x_{0}\right)\exp\left\{-\frac{\left[X+\sqrt{2}k_{2}^{3}\tau_{2}^{2}+i2\sqrt{2} k_{2}^{3}\tau_{2}\left(1+i\tau_{2}\right)\right]^{2}}{4\left[1+i\left(\tau_{2}+\tau_{3}\right)\right]}\right\}\right .\\
& \quad\pm\left .\exp\left(i\sqrt{2} k_{2}^{3}\tau_{2}x_{0}\right)\exp\left\{-\frac{\left[X-\sqrt{2}k_{2}^{3}\tau_{2}^{2}-i2\sqrt{2} k_{2}^{3}\tau_{2}\left(1+i\tau_{2}\right)\right]^{2}}{4\left[1+i\left(\tau_{2}+\tau_{3}\right)\right]}\right\}\right),
\end{split}
\end{equation}
\end{widetext}
with $M$ a normalization factor. Eq.(\ref{psimasmenosA}) is dimensionless thanks to the following definitions
\begin{align}\label{frankyA}
\tau_{2,3}& =\frac{\hbar t_{2,3}}{2m\sigma_{0}^{2}},&
   k_{2,3}& =\sqrt{2}\sigma_{0}\left(\frac{m\mu_{c}b_{2,3}}{2\hbar^{2}}\right)^{1/3},\nonumber\\
x_{0}& =\frac{B_{2}}{\sigma_{0}b_{2}},&
   z_{0}& =\frac{B_{3}}{\sigma_{0}b_{3}},\\
Z& =\frac{z}{\sigma_{0}},&
   X& =\frac{x}{\sigma_{0}}.\nonumber
\end{align}
The definitions in Eq. (\ref{frankyA}) make all the variables dimesionless and our results completely comparable with those of other works that study the entanglement in the SGE \cite{roston, Lenanuevo}. 
We found that the final state is a superposition state of the spin eigenstates, in the same way as the state coming out from a single SGE \cite{Ern,corrige}, and therefore does not represent a state following a definite trajectory to the screen. This state presents hybrid entanglement between the position DoFs and the spin DoF, congruently with the known effect present in the SGE, the separation of spins. The presence of the $Z$ variable in this entangled state attests the non-commutativity between the different spin operators for the different spatial orientations, this is in good accordance with the observations of the semiclassical argument for the thought experiment. We also found an interesting effect for the configuration proposed here in the presence of entanglement also with the position DoF in the $X$ coordinate.




\section{Steering}
\label{sec:Steering}

As was mentioned earlier, steering is a manifestation of quantum correlations. It was defined by Schr\"odinger (as a property of entangled systems) as the possibility that by suitable measurements taken on one subsystem only, the state of the other subsystem can be determined by the choice of the measurement and without interacting with it. That is to say, it is possible to steer the sate of a subsystem by choosing what kind of measurement to implement on the other subsystem.

The actual understanding of steering comes from the operational definition given by Wiseman et.al. \cite{wiseman2007}, they define steering as a task, i.e. the task of Alice is to convince Bob that she can prepare a bipartite entangled state. To do it, Alice prepares a bipartite quantum state and sends one of them to Bob; then, they measure their respective subsystem. If the correlations between their measurement can be explained by a local hidden state model for Bob, then Alice could have taken a pure state at random and sent it to Bob. But if the measurements can not be explained by a local hidden state model, then Alice steers Bob's state. Wiseman et.al. \cite{wiseman2007} demonstrated that steering is stronger than nonseparability and weaker than nonlocality. The existence of nonlocality rules out local hidden variables and, in essence, steering rules out the existence of local hidden state models  \cite{wiseman2007}. Steering was experimentally proven in single photon experiment by Guerreiro et al. \cite{brunner16}. In the case of pure tripartite states, He and Reid have demonstrated that it is enough that each party can be steered by one or both of the other two to certify steering \cite{he13}.

In reference \cite{are20}, it was demonstrated that in the case of a single SGE Alice can steer Bob's state depending on the measurement she chooses. In that paper \cite{are20}, a thought experiment where Alice is located in Tokyo and Bob in Paris was posed. Hence, by choosing which one of the possible observables to measure, Alice can steer Bob's state. This confirms the nonlocality of the entangled wavefunction of the SGE and reference \cite{are20} associates this nonlocality with the spreading of the entangled wavefunction.
\begin{figure}[h!]
  \centering
  \includegraphics[width=88mm]{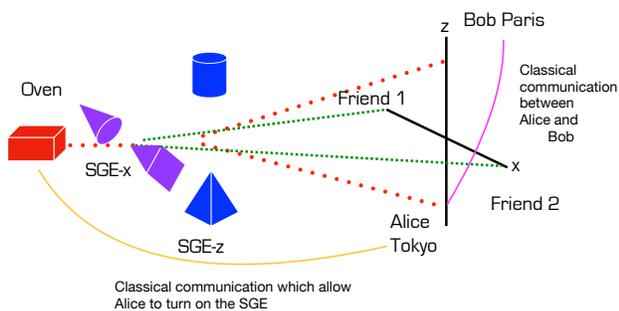}
  \caption{CSGE featuring the Einstein's boxes, see reference \cite{are20}. The red box is the oven, in violet a SGE in the $x$ direction is depicted, in blue the SGE in $z$ direction is depicted, the red and green dots represent the fact that there are not classical trajectories, see references \cite{Ern,corrige}. Alice could communicate with Bob by using the classical channel in magenta. Moreover, Alice is in full control of the SGE by using the classical channel in yellow, and she possess the ability to turn it on and to choose between a single or $N$ atoms. The red dots end in the $z$ axis in Tokyo with Alice and in Paris with Bob. The green dots in the $x$ axis arrives at suitable places where there are two friends of Alice.}
  \label{tokyo}
\end{figure}

In the case of the CSGE a similar situation --in fact more rich--  can be conceived. This is depicted in FIG (\ref{tokyo}), Alice in Tokyo is in full control of a fully automatized CSGE located at a suitable place and can turn it on and off and sent a single or many atoms one by one as she wishes.

Therefore, if Alice mades a check of $\ket{\downarrow_{z}}$ and she founds $-\hbar/2$, then the entangled wavefunction given by Eq. (\ref{3mod3}), i.e. 
$
\ket{\psi_{f}}=\ket{\psi_{+}}\ket{\uparrow_{z}}+\ket{\psi_{-}}\ket{\downarrow_{z}},\nonumber
$
would collapse to $\ket{\psi_{-}}\ket{\downarrow_{z}}$ in Tokyo, notice that $\ket{\psi_{-}}$ comprises a superposition state in $X$. On the other hand, if she detected nothing, then the wave function would collapse to $\ket{\psi_{+}}\ket{\uparrow_{z}}$ in Paris, see reference \cite{are20} for an explanation of this effect in terms of the Einstein's boxes, similarly, $\ket{\psi_{+}}$ comprises a superposition state in $X$.

However, if Alice decides to measure in a different basis, for example  $\hat{\sigma_x}$, to ascertain the possibilities we must rewritte Eq. (\ref{3mod3}) in the $\hat{\sigma_x}$ basis, getting
\begin{equation}\label{}
\ket{\psi_{f}}=\left\{\ket{\psi_{+}}+\ket{\psi_{-}}\right\}\ket{\uparrow_{x}}+\left\{\ket{\psi_{+}}-\ket{\psi_{-}}\right\}\ket{\downarrow_{x}},
\end{equation}
then, the following situations can arise:
\begin{enumerate}
\item[i)] If she measures spin down in the basis of $\hat{\sigma}_x$ and obtains $-\hbar/2$, then the wavefunction would collapse towards $\left\{\ket{\psi_{+}}-\ket{\psi_{-}}\right\}\ket{\downarrow_{x}}$. Notice that $\left\{\ket{\psi_{+}}-\ket{\psi_{-}}\right\}$ comprises a superposition state in $Z$.
\item[ii)] If she checks spin up in the basis of $\hat{\sigma}_x$ and obtains $\hbar/2$, then the wavefunction would collapse towards $\left\{\ket{\psi_{+}}+\ket{\psi_{-}}\right\}\ket{\uparrow_{x}}$.
\end{enumerate}
Therefore, we must conclude that Bob's state is steered depending on the kind of observable Alice decides to measure, confirming the nonlocality of the wavefunction generated by the CSGE. 

On the other hand, a similar procedure could be used by Bob in Paris to steer Alice's state. Or by one of the friends of Alice to steer states to Alice or Bob. Therefore the He and Reid criterion cited above is fulfilled, i.e.  the CSGE possesses the nonlocal property of steering. As it can be seen in FIG. (\ref{tokyo}) a richer situation that the one given in reference \cite{are20} arises, because in the CSGE case there could be the participation of four people to address all the possibilities. For example, if Alice's Friend 2 in a different location decides to check for $\hbar/2$ in the $\hat{\sigma_x}$ basis and obtains nothing in their measurement, then the wavefunction would collapse towards the state  $\left\{\ket{\psi_{+}}-\ket{\psi_{-}}\right\}\ket{\downarrow_{x}}$. This is equivalent to Alice measuring $\hat{\sigma_x}$ and obtaining $-\hbar/2$ and is also equivalent to Friend 1 measuring $\hat{\sigma_x}$ and obtaining $-\hbar/2$, both of them will collapse the wavefunction (by effects of their measurements) towards $\left\{\ket{\psi_{+}}-\ket{\psi_{-}}\right\}\ket{\downarrow_{x}}$, i.e. the same function as that obtained by Friend 2 when testing $\hbar/2$ in the basis $\hat{\sigma_x}$ and obtaining nothing.\par




\section{Non-locality of the Consecutive Stern Gerlach experiments}
\label{sec:nonloc}
The study of quantum correlations of quantum systems is strongly related with non-clasical tasks \cite{wolfe20,chaves17,araujo20,roston,Ern,corrige,potel,epic1,home22,Lenanuevo,Haug_2004,Ferraro_2005,Chen_2002,Chen2,bancal,BancalDef,Wodkiewicz,Banaszek_1999,wisin98} that open the way to important applications. We quantify the quantum correlations of the CSGE with a correlation function for hybrid spin systems that has already been tested for bipartite states \cite{Wodkiewicz,epic1} 
\begin{equation}\label{corrp}
\bm{C}(x,p_x,z,p_z,\theta)=\expval{\hat{W}(x,p_x, z, p_z)\hat{\sigma}(\theta)}{\psi_{f}},
\end{equation} 
that is the generalized Banaszek-W{\'{o}}dkiewicz (BW) correlation function \cite{Banaszek_1999}, with
\begin{equation}\label{papawigner}
\begin{split}
\hat{W}&(x,p_x, z, p_z) =\int_{-\infty}^{\infty} dq_{x} dq_{z}\ket{x-\frac{1}{2} q_{x}, z-\frac{1}{2} q_{z}}\\
& \times\exp[-i(p_{z}q_{z}+p_{x}q_{x})/\hbar]\bra{x+\frac{1}{2}q_{x}, z+\frac{1}{2}q_{z}},
\end{split}
\end{equation}
the generalized Wigner operator, originally defined by Ben-Benjamin et. al. \cite{Ben_Benjamin_2016}, for coordinates $x$, $z$ with their respective moments $p_{x}$, $p_{z}$; $\hat{\sigma}(\theta)$ is the usual Pauli operator for an arbitrary $\theta$ direction in the plane.\par
In this way we can calculate the correlation function for the final state of the CSGE that is given by Eq. (\ref{3mod3}).
 The relation between the correlations and entanglement in the CSGE will be discussed later in this paper.\par
\subsection{Results}
\label{subsec:Results}
Our quantum description of the experiments allows us to delve into the quantum characteristics of the system, in this manner, we study the quantum correlations that are present by means of the CHSH, Bell-Klyshko-Mermin and Svetlichny inequalities.  
\subsubsection{CHSH inequality}
From Eq. (\ref{corrp}) the correlation function of the CSGE has the following form:
\begin{widetext}
\begin{equation}\label{correlg}
\begin{split}
&\bm{C}(X,P_x,Z,P_z,\theta) =\exp[\omega^{\prime}_{z}(Z,P_z)+\omega^{\prime}_{x}(X,P_x)]\\
& \times\Bigg{[}4\cos(\theta)\bigg{(}\exp[\omega_{z}]\Big{\{}\exp[\omega_{x}]\cosh[d_{x}(X,P_x)]\sinh[d_{z}(Z,P_z)]\\
&\qquad+\exp[-\omega_{x}]\cos[\delta_{x}(X,P_x)]\cosh[d_{z}(Z,P_z)]\Big{\}}\bigg{)}\\
&\ \ \ +4 \sin(\theta)\bigg{(}\exp[-\omega_{z}]\Big{\{}\exp[\omega_{x}] \sinh[d_{x}(X,P_x)] \cos[\delta_{z}(Z,P_{z})]\\
&\qquad+\exp[-\omega_{x}]\sin[\delta_{x}(X,P_x)] \sin[\delta_{z}(Z,P_z)]\Big{\}}\bigg{)} \Bigg{]}
\end{split}
\end{equation} 
\end{widetext}
up to a normalization factor. The functions $\omega^{\prime}_{z}(Z,P_{z})$, $\omega^{\prime}_{x}(X,P_{x})$, $\omega_{z}$, $\omega_{x}$, $d_{z}(Z,P_{z})$, $d_{x}(X,P_{x})$, $\delta_{z}(Z,P_{z})$ and $\delta_{x}(X,P_{x})$ are real valued functions. See the Appendix B for the complete definitions.\par
We have the extra dimensionless definitions given by
\begin{align}\label{franky2}
P_{x}& =\frac{p_{x}\sigma_{0}}{\hbar},&
P_{z}& =\frac{p_{z}\sigma_{0}}{\hbar}.
\end{align}
The CHSH inequality \cite{Clauser1969,Ferraro_2005,wisin98,Wodkiewicz} that we use to test the existence of non-locality bet\-ween the $Z$-$\theta$ pair is as follows
\begin{equation}\label{chsczthe}
\begin{split}
-2\leq & B_{CHSH}=\bm{C}( X,P_x,Z,P_z,\theta)+\bm{C}( X,P_x,Z,P_z,\theta^{\prime})\\
& +\bm{C}( X,P_x,Z^{\prime},P_z,\theta)-\bm{C}( X,P_x,Z^{\prime},P_z,\theta^{\prime})\leq 2.
\end{split}
\end{equation}
To study the non-locality between $z$ and $\theta$ we need to fix the other variables, named $X$, $P_x$, $P_{z}$, $Z^{\prime}$ and $\theta^{\prime}$. \par
\begin{figure}[h!]
  \centering
  \includegraphics[width=86mm]{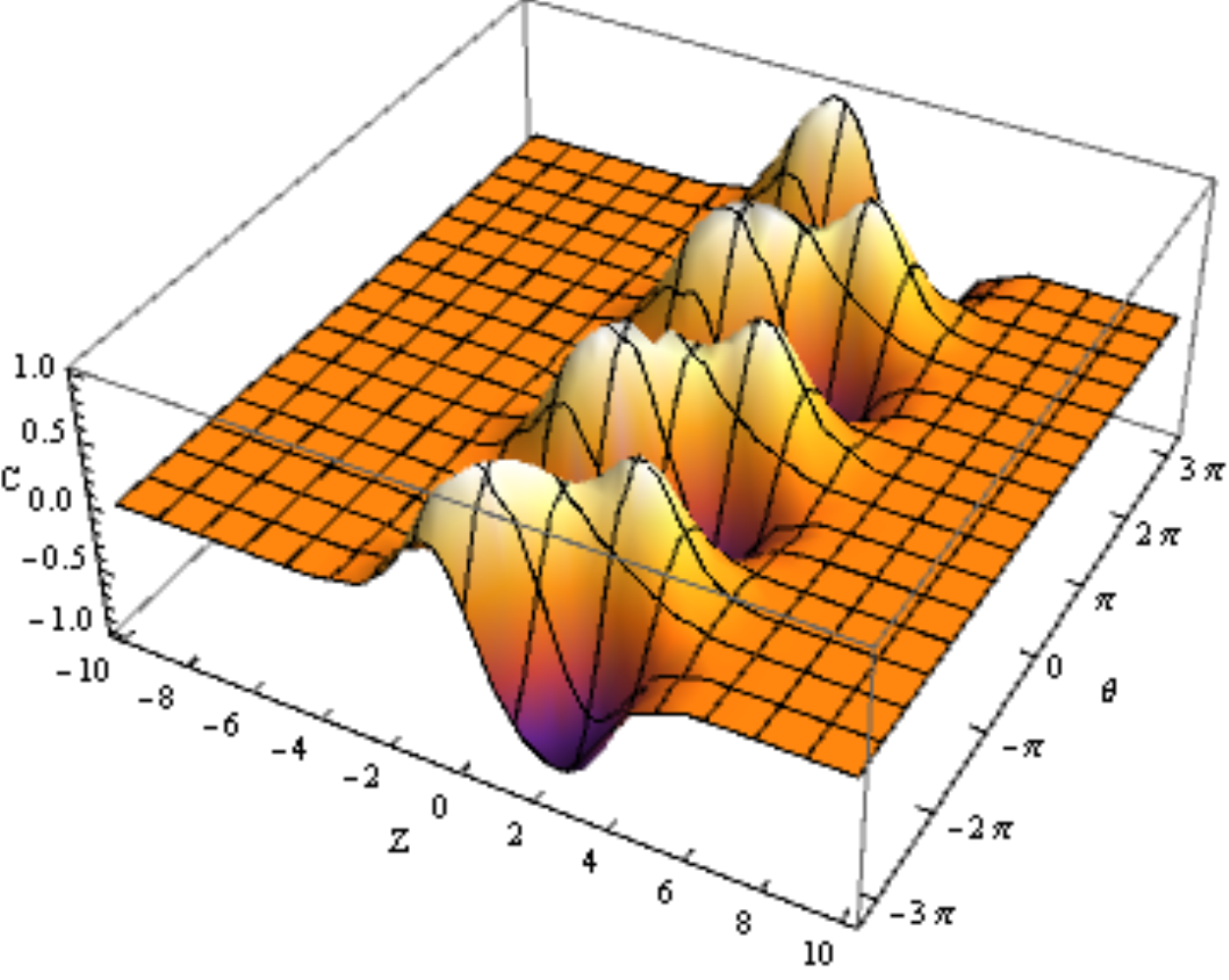}
  \caption{Correlation function for the $Z$-$\theta$ pair taking the fixed quantitites as $X=0.1$, $P_x=0.129$, $P_{z}=0.049$, $\tau_{2}=6.8$, $\tau_{3}=2.6$, $k_{2}=0.3$, $k_{3}=0.3$, $x_{0}=4$ and $z_{0}=4$.}
  \label{correl}
\end{figure}
\begin{figure}[h!]
  \centering
  \includegraphics[width=86mm]{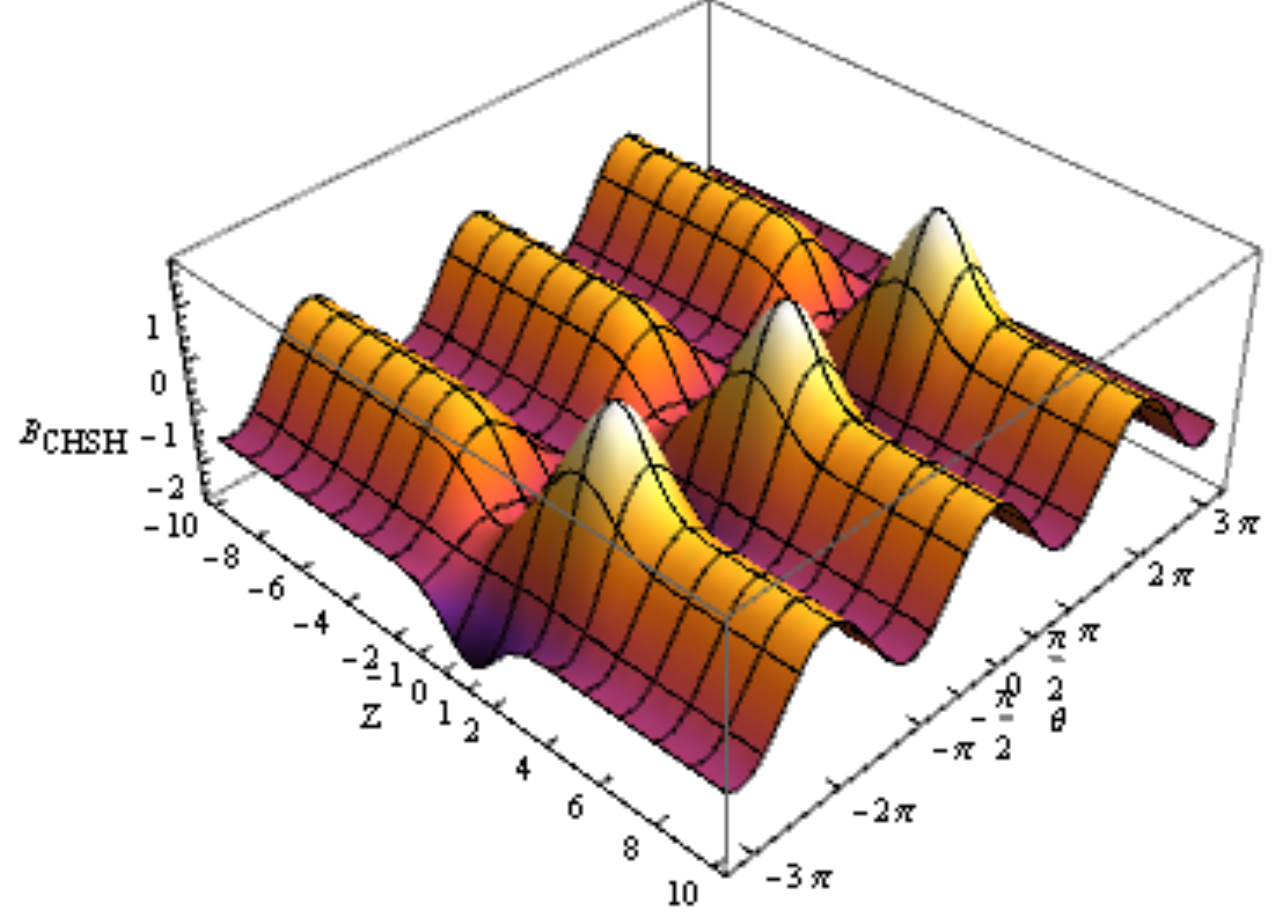}
  \caption{Plot of $B_{CHSH}$ for the primed quantities  $Z^{\prime}=2.4$ and $\theta^{\prime}=\frac{\pi}{5}$. We found a minimum of $-2.62405$ for this function, in this way, there exists a violation for the inequiality \eqref{chsczthe} by an amount of $0.62405$, i.e., around of $75\%$ of the maximal amount of violation, $\sim 0.8284$, given by the Cirel'son's bound\cite{Ferraro_2005,wisin98,cirelson}.}
  \label{bellzz}
\end{figure}

The plot of the correlation function for this case appears in FIG. \ref{correl} for the values of the fixed parameters $X=0.1$, $P_x=0.129$, $P_{z}=0.049$, $\tau_{2}=6.8$, $\tau_{3}=2.6$, $k_{2}=0.3$, $k_{3}=0.3$, $x_{0}=4$ and $z_{0}=4$. The Bell function for this case, $B_{CHSH}$, is shown in FIG. \ref{bellzz} for the primed quantities  $Z^{\prime}=2.4$ and $\theta^{\prime}=\frac{\pi}{5}$. We report the violation of the CHSH inequality for the $Z$,$\theta$ pair for a minimum value of  $B_{CHSH}$ of (negative) -2.62405.
\subsubsection{Bell-Klyshko-Mermin inequality}
Our treatment for the study of correlations and the number of DOFs in the system allows us to evaluate tripartite non-locality, the case of major interest is the $X$-$Z$-$\theta$ triad. The Bell-Klyshko-Mermin inequality that we use for this purpose is the following \cite{Ferraro_2005,wisin98}:
\begin{equation}\label{bkmxzthe}
\begin{split}
-2&\leq B_{BKM}= \bm{C}(X,P_x,Z,P_z,\theta^{\prime}) +\bm{C}(X,P_x,Z^{\prime},P_z,\theta)\\
& +\bm{C}(X^{\prime},P_x,Z,P_z,\theta)-\bm{C}(X^{\prime},P_x,Z^{\prime},P_z,\theta^{\prime})\leq 2.
\end{split}
\end{equation}
In this case we need to fix the $P_{X}$, $P_{Z}$, $X^{\prime}$, $Z^{\prime}$ and $\theta^{\prime}$ variables. 
The verification of  Eq. (\ref{bkmxzthe}) can only be handled numerically and it is not possible to visualize a plot of $B_{BKM}$ due to the number of DoFs in the Bell-Klyshko-Mermin inequality. Our study of the inequality of  Eq.\eqref{bkmxzthe} gives us as maximum value for $B_{BKM}$ the quantity $2.43258$, demostrating the violation of the inequality \eqref{bkmxzthe} for the fixed quantities $P_{x}=0.051$, $P_{z}=0.089$, $X^{\prime}=0.83$, $Z^{\prime}=3.3$ and $\theta^{\prime}=\frac{\pi}{2}$, $\tau_{2}=2.7$, $\tau_{3}=4.7$, $k_{2}=0.3$, $k_{3}=0.3$, $x_{0}=4$ and $z_{0}=4$.
\subsubsection{Svetlichny Inequality}
Our third test of non-locality is given by the Svetlichny inequality \cite{Svetlichny1987,lavoie,bancal} for $X$, $Z$ and $\theta$. The Svetlichny inequality is given as
\begin{widetext}
\begin{equation}\label{svet}
\begin{split}
-4& \leq B_{S}=\bm{C}(X,P_x,Z,P_z,\theta)+\bm{C}(X,P_x,Z,P_z,\theta^{\prime})\\
& \qquad +\bm{C}(X,P_x,Z^{\prime},P_z,\theta)-\bm{C}(X,P_x,Z^{\prime},P_z,\theta^{\prime})+\bm{C}(X^{\prime},P_x,Z,P_z,\theta)\\
& \qquad -\bm{C}(X^{\prime},P_x,Z,P_z,\theta^{\prime})-\bm{C}(X^{\prime},P_x,Z^{\prime},P_z,\theta)-\bm{C}(X^{\prime},P_x,Z^{\prime},P_z,\theta^{\prime})\leq 4,
\end{split}
\end{equation}
\end{widetext}
fixing $P_{x}$, $P_{z}$, $X^{\prime}$, $Z^{\prime}$ and $\theta^{\prime}$. Our exhaustive study of the inequality of Eq. (\ref{svet}) permits us to report a possible non-violation of the Svetlichny inequality for our system. We obtain the maximum of $B_{S}\lessapprox 4$ for the fixed quantities $P_{x}=0.043$, $P_{z}=0.066$, $X^{\prime}=0.52$, $Z^{\prime}=1.49$ and $\theta^{\prime}=\frac{\pi}{5}$, $\tau_{2}=2.3$, $\tau_{3}=3.5$, $k_{2}=0.3$, $k_{3}=0.3$, $x_{0}=4.5$ and $z_{0}=5.2$.\par
Nonetheless, we have tested our system with two weaker versions of the Svetlichny inequality given by reference \cite{bancal},
\begin{widetext}
\begin{equation}\label{svetweak1}
\begin{split}
B_{SV1}=& -\bm{C}(X,P_x,Z,P_z,\theta)+\bm{C}(X,P_x,Z^{\prime},P_z,\theta)\\
& +\bm{C}(X^{\prime},P_x,Z,P_z,\theta)+\bm{C}(X^{\prime},P_x,Z^{\prime},P_z,\theta)+\bm{C}(X,P_x,Z,P_z,\theta^{\prime})\\
& +\bm{C}(X,P_x,Z^{\prime},P_z,\theta^{\prime})+\bm{C}(X^{\prime},P_x,Z,P_z,\theta^{\prime})-\bm{C}(X^{\prime},P_x,Z^{\prime},P_z,\theta^{\prime})\leq 4,
\end{split}
\end{equation}
\end{widetext}
and by inequality $185$ in reference \cite{BancalDef},

\begin{widetext}
\begin{equation}\label{svetweak2}
\begin{split}
B_{SV2}=& -\bm{C}(X,P_x,Z,P_z,\theta)-\bm{C}(X,P_x,Z^{\prime},P_z,\theta)\\
& +\bm{C}(X^{\prime},P_x,Z,P_z,\theta)-\bm{C}(X^{\prime},P_x,Z^{\prime},P_z,\theta)-\bm{C}(X,P_x,Z,P_z,\theta^{\prime})\\
& +\bm{C}(X,P_x,Z^{\prime},P_z,\theta^{\prime})-\bm{C}(X^{\prime},P_x,Z,P_z,\theta^{\prime})-\bm{C}(X^{\prime},P_x,Z^{\prime},P_z,\theta^{\prime})\leq 4.
\end{split}
\end{equation}
\end{widetext}

In both cases, for $B_{SV1}$ and $B_{SV2}$, we have not found any violation for a wide variety of cases with different variables; after a wide search for conditions to find a possible violation, this suggest a posible no-violation of this quantities. This result is in concordance with the maximum found in the Bell-Klyshko-Mermin inequality, Eq. \eqref{bkmxzthe}, because the maximal violation is not greater than $2\sqrt{2}$ for our case \cite{wisin98, cereceda}. Additionally, it is convenient to recall that in certain cases stronger nonlocal effects are due to weaker entangled states \cite{vidick,dilley18,scarani07}, in such cases the maximun entangled states do not produce the maximun nonlocality \cite{scarani07}.


\section{Relation between correlations and entanglement in the CSGE}
\label{sec:entang}
The entanglement in the SGE has been widely studied in recent years, for instance in \cite{roston,home22,Lenanuevo}. One of the primal goals of these descriptions of entanglement is the quantification of the same by means of entanglement measures either in discrete, continous or hybrid sytems.\par
Our treatment of the quantum correlations is directly equiparable with the study of the entanglement of the SGE by \cite{roston,Lenanuevo} due to it maintaining the same quantum description of the SGE of \cite{Lenanuevo} and the same temporal dependence in the correlation function, Eq. (\ref{correlg}), and the entanglement of \cite{Lenanuevo,roston}.\par
\begin{figure}[h!]
  \centering
  \includegraphics[width=86mm]{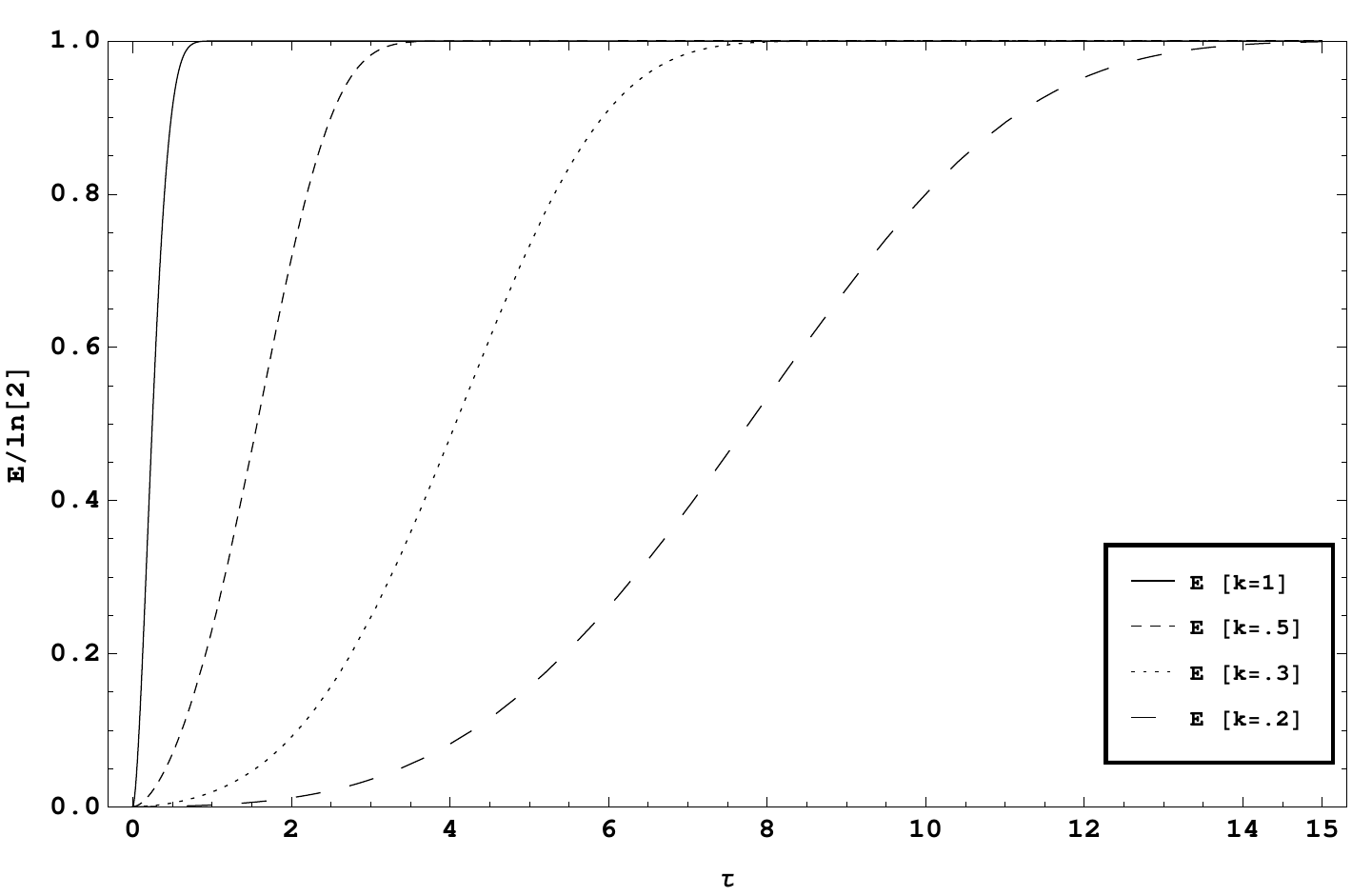}
  \caption{Entanglement entropy ($E$) of the SGE for various values of $k$ trough the time $\tau$. In the CSGE, we have $k_{2,3}$ and $\tau_{2,3}$, which are equivalent to these non-labeled constants. In this way we can directly compare the temporal behavior of the quantum correlations and entanglement in the CSGE relying in the parameter $k$. Plot similar to the one given in \cite{Lenanuevo}.}
  \label{entangle}
\end{figure}

For the case of the CHSH inequality, the maximum violation found in the inequality of  Eq. (\ref{chsczthe}) is reached for $\tau_{2}=6.8$, $\tau_{3}=2.6$, $k_{2}=0.3$, $k_{3}=0.3$. For values of the addimensional time $\tau_{2}=6.8$ with $k_{2}=0.3$, the maximum entanglement between $X$ and $\theta$ is almost achieved after the first SGE, while for $\tau_{3}=2.6$ with $k_{3}=0.3$ in a single SGE, the entanglement would be small; see FIG. \ref{entangle}. Therefore, in this case, to get the maximum quantum nonlocality the atom does not need to expend much time in the second SGE for the CSGE,
 the only thing that we can assure is that in the presence of maximal quantum correlations exists entanglement. \par
We report a non-violation of the CHSH inequality of the CSGE for times, $\tau_{2,3}$ where the entanglement is maximum, for example $\tau_{2,3}=10$ for $k_{2,3}=0.3$, a counterintuitive result at first glance but that has its explanation in the dishtinguishablity that the SGE state presents when time tends to infinity, a fact that is endorsed by the probability of the final state of the SGE \cite{Ern,corrige}, wich also could be calculated for the CSGE final state of Eq. (\ref{3mod}).\par
We expect that for the case with more DoFs engaged, the entanglement in the CSGE will have a similar behavior as the original SGE has, this is a reasonable expectation. In this manner, we also report a non-violation of the Bell-Klyshko-Mermin and the weaker Svetlichny inequalities when $\tau_{2,3}\rightarrow\infty$, meaning that, in these cases, we anticipate the lack of quantum correlations when the multipartite entanglement on those cases will be maximum.
\section{Conclusions}
\label{sec:Conclusions}
In this work we have studied the non-locality and the steering in the CSGE. We have demonstrated that the CSGE could be used to steer quantum states between two different places. Our analysis suggest that the spreading of the entangled wave function allows the particle to sense which observable is being measured by Alice; hence, suggesting that the particle senses which observable is being measured, for example $\hat{\sigma}_x$, $\hat{\sigma}_z$ or $\hat{x}$.

Additionally, we have found violation of the CHSH and Bell-Klyshko-Mermin inequalities, however we have not found data suggesting that the evolved state of the CSGE violate the strong Svetlichny inequality nor the two weaker forms of the Svetlichny inequality. We then conclude the presence of non-locality in the CSGE in the bipartite case but we can not assure the existence of tripartite  nonlocality nor real tripartite entanglement due to the lack of violation of the Svetlichny inequalities \cite{wisin98,cereceda,lavoie}. This result does not imply that a violation of the Svetlichny inequality can not exist for the CSGE just that we can not provide a proof of violation of such inequality, further research is necessary to find out whether of not it is violated by this state.\par

The striking result of the SGE with the spatial separation of the spins is carried to the CSGE evolution, where our description of the evolved state indicates the presence of entanglement between spin and position. The quantum character of the SGE is then raised to attention when proving the correlations between position and spin to be non-local, and therefore, non-classical in nature.
The quantum correlations of the hybrid tripartite state of the CSGE were successfully characterized by a generalization of the BW correlation function. The proposed function dichotomizes the spatial DoFs using the parity operator and combines this description with the usual one for discrete variables \cite{Wodkiewicz}. With these results, the CSGE presents an important case of hybrid non-locality.\par

Furthermore, the violation of the CHSH inequality by the correlated $Z$-$\theta$ pair, $Z$ being the direction of the last separation by the SGE apparatus, was found to have a maximum value of $2.62$. This is a sizeable, considerable violation, appreciably close to the maximum value for the violation of the two-party CHSH inequality of $2\sqrt{2}$. Also, a very good violation for this type of Bell operators \cite{Chen_2002,Chen2}.\par

On the other hand, the maximal violation of the tripartite Bell-Klyshko-Mermin inequality found for the CSGE state is 2.43 for the $X$-Z-$\theta$ triad. This proves tripartite non-locality in the CSGE, but this inequality detects as well the non-locality of the multiple two-party correlations. The violation found is in this case far away from the possible maximum of 4 for the tripartite inequality \cite{wisin98}.\par
We can not establish the presence of genuine tripartite (three-way) non-locality from the lack of violation of the Svetlichny inequalities, moreover, real hybrid tripartite entanglement is also not assured in our system \cite{bancal,BancalDef,cereceda}. \par

We can further correlate these results with the presence of entanglement in the SGE. In the usual SGE, hybrid entanglement is created as soon as we have interaction with the inhomogeneous field, with it increasing with the time of evolution in the field until reaching the maximum possible entanglement \cite{roston,Lenanuevo}. At this point, the correlations between position and spin are perfect. We can reasonably expect these results to be carried over to the CSGE. However, in the whole configuration space two-party non-locality has not been found for parameters of the CSGE evolution where we expect maximal entanglement. Tripartite non-locality in the form of a violation of the Bell-Klyshko-Mermin inequality is also not found for the parameters that would give maximal entanglement. This result is in concordance with previous studies where it is shown that systems with maximum entanglement present slightly or null nonlocality.\par
We can ensure that in the points where non-locality has been found, there exists tripartite entanglement in the state of the CSGE, but this entanglement is not perfect.\par
\ \par
E. Ben\'itez Rodr\'iguez and E. Piceno Mart\'inez thank CONACYT for PhD fellowship support. We thank Vicerrector\'ia de Investigaci\'on y Estudios de Posgrado (BUAP) for partial support. This work has not been supported by any scientific agency.




\appendix

\section{Dynamic evolution of the CSGE}
\label{sec:DynamicSM}
In this part we will find the explicit evolution of the initial wavepacket as it traverses or configuration of CSGE of FIG. 1.\par
The effective Hamiltonian related with the first ex\-pe\-ri\-ment is \cite{Ern,corrige}
\begin{equation}
\label{H2a}
\hat{H}_2=-\frac{\hbar^2}{2m}\nabla^2+\mu_\text{c}(\bm{\sigma}\cdot\bm{B}_2),
\end{equation}
with the effective inhomogeneous magnetic field $\bm{B}_2=(B_2+b_2x)\hat{\imath}$, $\mu_{c}=ge\hbar/(4m)$, $g$ the gyromagnetic ratio, $m$ the mass of the particle, $e$ the unit charge and $\bm{\sigma}$ the Pauli matrices vector. In a similar way we have the effective Hamiltonian associated with the second experiment
\begin{equation}
\label{H3a}
\hat{H}_3=-\frac{\hbar^2}{2m}\nabla^2+\mu_{\text{c}}(\boldsymbol{\sigma}\cdot\boldsymbol{B}_3)
\end{equation}
with the effective field $\boldsymbol{B}_3=(B_3+b_3z)\hat{k}$.\par
The initial state is given in Eq. \eqref{preparado} of the article as
\begin{equation}\label{prepaAa}
\ket{\psi_{i}}=\frac{1}{(2\pi\sigma_{0}^{2})^{\frac{3}{4}}}\exp\left(-\frac{(x^{2}+y^{2}+z^{2})}{4\sigma_{0}^{2}}+ik_{y}y\right)\ket{\uparrow_z}.
\end{equation}
To obtain the evolution through the first experiment we have the evolution operator associated with the Hamiltonian $\hat{H_{2}}$ of Eq. (\ref{H2a}), appearing in Eq. \eqref{u2r} of the article
\begin{equation}\label{U22a}
\begin{split}
\hat{U}_2(t)& =\exp\left(-\frac{1}{6}\kappa\right)
	\exp\left[-\frac{i t}{2m\hbar}\left(p_y^2+p_z^2\right)\right]\\
	& \times\exp\left[-\frac{i t\mu_c}{\hbar}\left(B_2+b_2x\right)\sigma_x\right]
\exp\left(\frac{i t^2\mu_cb_2}{2m\hbar}p_x\sigma_x\right)\\
& \times\exp\left(-\frac{i t}{2m\hbar}p_x^2\right),
\end{split}
\end{equation}
with $\kappa=(i t^2\mu_{c}^{2}b_{2}^{2})/(m\hbar)$, $B_{2}$ the homogeneity parameter of the magnetic field, $b_{2}$ the inhomogeneity parameter, $\sigma_{x}$ the Pauli operator. The evolution operator of this Eq. (\ref{U22a}) is factorized utilizing the Evolution Operator Factorization Method of \cite{are3,are4}, as done in \cite{Ern,corrige}.\par

Applying  $\hat{U}_2(t)$ for a certain time $t_{2}$ to $\ket{\psi_{i}}$ we have the intermediate state $\ket{\psi_m}$,
\begin{widetext}
\begin{equation}\label{2moda}
\begin{split}
\hat{U}_2(t_2)\ket{\psi_{i}}&=\ket{\psi_{m}(t_{2})}\\
& =\frac{1}{\sqrt{2}}\exp\left(-\frac{1}{6}\kappa_{2}\right)\sigma_{0}^{\frac{3}{2}}\left(\sigma_{0}^{2}+\frac{i\hbar t_{2}}{2m}\right)^{-\frac{3}{4}}\left\{\left[(2\pi)^{\frac{1}{2}}\left(\sigma_{0}^{2}+\frac{\mathrm{i\hbar t_{2}}}{2m}\right)^{\frac{1}{2}}\right]^{-\frac{1}{2}}\right\}^{3}\\
& \quad\times \exp(-k_{y}^{2}\sigma_{0}^{2})\exp\left\{\frac{-[(y-2ik_{y}\sigma_{0}^{2})^{2}+z^{2}]}{4(\sigma_{0}^{2}+\frac{i\hbar t_{2}}{2m})}\right\}\\
	&\quad \times\left\{\exp\left[-\frac{it_{2}\mu_{c}}{\hbar}(B_{2}+b_{2}x)\right]\exp\left[\frac{-(x+\frac{t^{2}_{2}\mu_{c}b_{2}}{2m})^{2}}{4(\sigma_{0}^{2}+\frac{i\hbar t_{2}}{2m})}\right]\ket{\uparrow_{x}}\right .\\
&\qquad +\left .\exp\left[\frac{it_{2}\mu_{c}}{\hbar}(B_{2}+b_{2}x)\right]\exp\left[\frac{-(x-\frac{t^{2}_{2}\mu_{c}b_{2}}{2m})^{2}}{4(\sigma_{0}^{2}+\frac{i\hbar t_{2}}{2m})}\right]\ket{\downarrow_{x}}\right\}
\end{split}
\end{equation}
\end{widetext}
where we remember $\ket{\uparrow_{z}}=\frac{1}{\sqrt{2}}(\ket{\uparrow_{x}}+\ket{\downarrow_{x}})$, and $\kappa_2=\kappa(t_2)$. \par
Now, following our description of the CSGE, we apply the evolution operator associated with the last SGE for a fixed time $t_{3}$. That is, we apply the evolution operator in direction $z$ direction, corresponding to the Hamiltonian of Eq. (\ref{H3a}), to the state of the Eq. (\ref{2moda}). This operator is given by

\begin{equation}\label{u3ra}
\begin{split}
\hat{U}_3(t)&=\exp\left(-\frac{1}{6}\kappa\right)
	\exp\left[-\frac{i t}{2m\hbar}\left(p_x^2+p_y^2\right)\right]\\
& \times\exp\left[-\frac{i t\mu_c}{\hbar}\left(B_3+b_{3}z\right)\sigma_z\right]
	\exp\left(\frac{i t^2\mu_cb}{2m\hbar}p_z\sigma_z\right)\\
& \times \exp\left(-\frac{i t}{2m\hbar}p_z^2\right)
\end{split}
\end{equation}
Then we obtain the final state of the system, $\ket{\psi_f}$,
\begin{widetext}
\begin{equation}\label{3modaa}
\begin{split}
\hat{U}_{3}(t_3)&\ket{\psi_{m}( t_{2})} =\ket{\psi_{f}} \\
& =A'_{3}\times\exp\left\{\frac{-(y-2ik_{y}\sigma_{0}^{2})^{2}}{4\left[\sigma_{0}^{2}+\frac{i\hbar( t_{2}+ t_{3})}{2m}\right]}\right\}\\
& \times\left[\exp\left[-\frac{i t_{3}\mu_{c}}{\hbar}(B_{3}+b_{3}z)\right]\exp\left\{\frac{-\left(z+\frac{ t^{2}_{3}\mu_{c}b_{3}}{2m}\right)^{2}}{4[\sigma_{0}^{2}+\frac{i\hbar( t_{2}+ t_{3})}{2m}]}\right\}\right .\\
& \times\left(\exp\left\{-\frac{i t_{2}\mu_{c}}{\hbar}B_{2}\right\}\exp\left\{\frac{-\left[x+\frac{ t^{2}_{2}\mu_{c}b_{2}}{2m}+2i\frac{ t_{2}\mu_{c}b_{2}}{\hbar}(\sigma_{0}^{2}+\frac{i\hbar t_{2}}{2m})\right]^{2}}{4[\sigma_{0}^{2}+\frac{i\hbar( t_{2}+ t_{3})}{2m}]}\right\}\right .\\
& \quad+\left .\exp\left\{\frac{i t_{2}\mu_{c}}{\hbar}B_{2}\right\}\exp\left\{\frac{-\left[x-\frac{ t^{2}_{2}\mu_{c}b_{2}}{2m}-2i\frac{ t_{2}\mu_{c}b_{2}}{\hbar}(\sigma_{0}^{2}+\frac{i\hbar t_{2}}{2m})\right]^{2}}{4[\sigma_{0}^{2}+\frac{i\hbar( t_{2}+ t_{3})}{2m}]}\right\}\right)\ket{\uparrow_{z}}\\
		& +\left .\exp\left[\frac{i t_{3}\mu_{c}}{\hbar}(B_{3}+b_{3}z)\right]\exp\left\{\frac{-\left(z-\frac{ t^{2}_{3}\mu_{c}b_{3}}{2m}\right)^{2}}{4[\sigma_{0}^{2}+\frac{i\hbar( t_{2}+ t_{3})}{2m}]}\right\}\right .\\
& \times\left(\exp\left\{-\frac{i t_{2}\mu_{c}}{\hbar}B_{2}\right\}\exp\left\{\frac{-\left[x+\frac{ t^{2}_{2}\mu_{c}b_{2}}{2m}+2i\frac{ t_{2}\mu_{c}b_{2}}{\hbar}(\sigma_{0}^{2}+\frac{i\hbar t_{2}}{2m})\right]^{2}}{4[\sigma_{0}^{2}+\frac{i\hbar( t_{2}+ t_{3})}{2m}]}\right\}\right .\\
& \quad- \left .\left .\exp\left\{\frac{i t_{2}\mu_{c}}{\hbar}B_{2}\right\}\exp\left\{\frac{-\left[x-\frac{ t^{2}_{2}\mu_{c}b_{2}}{2m}-2i\frac{ t_{2}\mu_{c}b_{2}}{\hbar}(\sigma_{0}^{2}+\frac{i\hbar t_{2}}{2m})\right]^{2}}{4[\sigma_{0}^{2}+\frac{i\hbar( t_{2}+ t_{3})}{2m}]}\right\}\right)\ket{\downarrow_{z}}\right]\\
& \\
\end{split}
\end{equation}
\end{widetext}
with similarly $\kappa_3=\kappa(t_3)$ and
\begin{equation}\label{a3pa}
\begin{split}
A'_{3}& =\exp\left(\frac{-\kappa_{2}-\kappa_{3}}{6}\right)\frac{1}{2}\left[\frac{\sigma_{0}}{(2\pi)^{\frac{1}{2}}}\right]^{\frac{3}{2}}\\
& \times\left(\sigma_{0}^{2}+\frac{ i\hbar}{2m}( t_{2}+ t_{3})\right)^{-\frac{3}{2}}\\
& \times\exp\left(-k_{y}^{2}\sigma_{0}^{2}\right)\exp\left(\frac{ i t_{2}}{\hbar}\mu_{c}b_{2}\cdot\frac{ t_{2}^{2}\mu_{c}b_{2}}{2m}\right)\\
& \times\exp\left[-\left(\frac{ t_{2}\mu_{c}b_{2}}{\hbar}\right)^{2}\left(\sigma_{0}^{2}+\frac{i\hbar t_{2}}{2m}\right)\right]
\end{split}
\end{equation}\par
To obtain Eq. \eqref{3mod3} in the article, we rewrite the final state of Eq. (\ref{3modaa}) as follows

\begin{equation}
\ket{\psi_{f}}=\ket{\psi_{+}}\ket{\uparrow_{z}}+\ket{\psi_{-}}\ket{\downarrow_{z}},
\end{equation}
defining, as in Eq. \eqref{psimasmenosA} of the article,
\begin{widetext}
\begin{equation}\label{psimasmenos}
\begin{split}
\ket{\psi_{\pm}}& =M\exp\left(\mp i\sqrt{2}k_{3}^{2}\tau_{3}z_{0}\right)\exp\left\{-\frac{\left(Z\pm\left\{\sqrt{2}k_{3}^{3}\tau_{3}^{2}+i2\sqrt{2}k_{3}^{3}\tau_{3}\left[1+i\left(\tau_{2}+\tau_{3}\right)\right]\right\}\right)^{2}}{4\left[1+i\left(\tau_{2}+\tau_{3}\right)\right]}\right\}\\
& \times\left(\exp\left(-i\sqrt{2} k_{2}^{3}\tau_{2}x_{0}\right)\exp\left\{-\frac{\left[X+\sqrt{2}k_{2}^{3}\tau_{2}^{2}+i2\sqrt{2} k_{2}^{3}\tau_{2}\left(1+i\tau_{2}\right)\right]^{2}}{4\left[1+i\left(\tau_{2}+\tau_{3}\right)\right]}\right\}\right .\\
& \quad\pm\left .\exp\left(i\sqrt{2} k_{2}^{3}\tau_{2}x_{0}\right)\exp\left\{-\frac{\left[X-\sqrt{2}k_{2}^{3}\tau_{2}^{2}-i2\sqrt{2} k_{2}^{3}\tau_{2}\left(1+i\tau_{2}\right)\right]^{2}}{4\left[1+i\left(\tau_{2}+\tau_{3}\right)\right]}\right\}\right),
\end{split}
\end{equation}
\end{widetext}

with $M$ the normalization factor given from Eqs. (\ref{2moda}) and (\ref{a3pa}) by
\begin{widetext}
\begin{equation}
\begin{split}
M& =\exp\left(\frac{-\kappa_{2}-\kappa_{3}}{6}\right)\frac{1}{2}\left[\frac{\sigma_{0}}{(2\pi)^{\frac{1}{2}}}\right]^{\frac{3}{2}}\left(\sigma_{0}^{2}+\frac{\mathrm{i}\hbar}{2m}(\tau_{2}+\tau_{3})\right)^{-\frac{3}{2}}\\
& \times\exp\left(-k_{y}^{2}\sigma_{0}^{2}\right)\times\exp\left[-\left(\frac{ t_{2}\mu_{c}b_{2}}{\hbar}\right)^{2}\sigma_{0}^{2}\right]\\
& \times\exp\left[-\left(\frac{ t_{3}\mu_{c}b_{3}}{\hbar}\right)^{2}\left(\sigma_{0}^{2}+\frac{i\hbar t_{2}}{2m}\right)\right]\exp\left\{\frac{-(y-2ik_{y}\sigma_{0}^{2})^{2}}{4\left[\sigma_{0}^{2}+\frac{i\hbar( t_{2}+ t_{3})}{2m}\right]}\right\},\end{split}
\end{equation}
\end{widetext}
and the dimensionless definitions of Eq. \eqref{frankyA} in the article
\begin{align}\label{frankya}
\tau_{2,3}& =\frac{\hbar t_{2,3}}{2m\sigma_{0}^{2}},&
   k_{2,3}& =\sqrt{2}\sigma_{0}\left(\frac{m\mu_{c}b_{2,3}}{2\hbar^{2}}\right)^{1/3},\nonumber\\
x_{0}& =\frac{B_{2}}{\sigma_{0}b_{2}},&
   z_{0}& =\frac{B_{3}}{\sigma_{0}b_{3}},\\
Z& =\frac{z}{\sigma_{0}},&
   X& =\frac{x}{\sigma_{0}}.\nonumber
\end{align}

\section{The correlation function for the CSGE}
\label{sec:corric}
In the discussion surrounding Eq. (9) of the article, we found that the correlation function for the CSGE can be written as
\begin{widetext}
\begin{equation}\label{correlga}
\begin{split}
&\bm{C}(X,P_x,Z,P_z,\theta) =\exp[\omega^{\prime}_{z}(Z,P_z)+\omega^{\prime}_{x}(X,P_x)]\\
& \times\Bigg{[}4\cos(\theta)\bigg{(}\exp[\omega_{z}]\Big{\{}\exp[\omega_{x}]\cosh[d_{x}(X,P_x)]\sinh[d_{z}(Z,P_z)]\\
&\qquad+\exp[-\omega_{x}]\cos[\delta_{x}(X,P_x)]\cosh[d_{z}(Z,P_z)]\Big{\}}\bigg{)}\\
&\ \ \ +4 \sin(\theta)\bigg{(}\exp[-\omega_{z}]\Big{\{}\exp[\omega_{x}] \sinh[d_{x}(X,P_x)] \cos[\delta_{z}(Z,P_{z})]\\
&\qquad+\exp[-\omega_{x}]\sin[\delta_{x}(X,P_x)] \sin[\delta_{z}(Z,P_z)]\Big{\}}\bigg{)} \Bigg{]}
\end{split}
\end{equation} 
\end{widetext}
up to a normalization factor. 

We do so by defining the following functions that appear in Eq.\eqref{correlga}:

\begin{widetext}
\begin{equation}\label{wpza}
\begin{split}
&\omega'_{z}(Z,P_{z}) =\frac{1}{4\left[1+\left(\tau_{2}+\tau_{3}\right)^{2}\right]}\left[-\left(\tau_{2}+\tau_{3}\right)^{2}\left(Z^{2}+2k_{3}^{6}\tau_{3}^{4}\right)\right]-2k_{3}^{6}\tau_{3}^{2}\left[1+\left(\tau_{2}+\tau_{3}\right)^{2}\right]\\
& +2k_{3}^{6}\tau_{3}^{3}\left(\tau_{2}+\tau_{3}\right)-2\left[1+\left(\tau_{2}+\tau_{3}\right)^{2}\right]P_{z}^{2}-\frac{1}{4}Z^{2}+2P_{z}Z\left(\tau_{2}+\tau_{3}\right)\\
& -\frac{1}{4\left[1+\left(\tau_{2}+\tau_{3}\right)^{2}\right]}\left(Z^{2}+2k_{3}^{6}\tau_{3}^{4}\right)-4k_{3}^{6}\tau_{3}^{2}
\end{split}
\end{equation}
\end{widetext}

\begin{widetext}
\begin{equation}\label{wpxa}
\begin{split}
&\omega'_{x}(X,P_{x}) =\frac{1}{4\left[1+\left(\tau_{2}+\tau_{3}\right)^{2}\right]}\Big(-\left(\tau_{2}+\tau_{3}\right)^{2}\left(X^{2}+2k_{2}^{6}\tau_{2}^{4}\right)\\
&-8k_{2}^{6}\tau_{2}^{2}\left[4\left(1+\tau_{2}^{2}\right)^{2}-2\tau_{3}^{2}\left(1-\tau_{2}^{2}\right)+4\tau_{2}\tau_{3}\left(1+\tau_{2}^{2}\right)\right]\\
&+8k_{2}^{6}\tau_{2}^{3}\left\{4\left(\tau_{2}+\tau_{3}\right)-2\tau_{2}\left[1-\left(\tau_{2}+\tau_{3}\right)^{2}\right]\right\}\Big)\\
&-2\left[1+\left(\tau_{2}+\tau_{3}\right)^{2}\right]P_{x}^{2}-\frac{1}{4}X^{2}+2X P_{x}\left(\tau_{2}+\tau_{3}\right)\\
& -\frac{1}{4\left[1+\left(\tau_{2}+\tau_{3}\right)^{2}\right]}\Big{\{}X^{2}+2k_{2}^{6}\tau_{2}^{4} -8k_{2}^{6}\tau_{2}^{2}\left[2\left(1+\tau_{2}^{2}\right)+4\tau_{2}\tau_{3}\right]16k_{2}^{3}\tau_{2}^{3}\tau_{3}\Big{\}}
\end{split}
\end{equation}
\end{widetext}
\begin{widetext}
\begin{equation}\label{wza}
\begin{split}
\omega_{z}& =-\frac{1}{2}k_{3}^{6}\tau_{3}^{4}-2k_{3}^{6}\tau_{3}^{2}\left[1+\left(\tau_{2}+\tau_{3}\right)^{2}\right]
+2\left(\tau_{2}+\tau_{3}\right)k_{3}^{6}\tau_{3}^{3}
\end{split}
\end{equation}
\end{widetext}

\begin{widetext}
\begin{equation}\label{wxa}
\begin{split}
\omega_{x}=-\frac{1}{2}k_{2}^{6}\tau_{2}^{4}-2k_{2}^{6}\tau_{2}^{2}
\end{split}
\end{equation}
\end{widetext}

\begin{equation}\label{dza}
\begin{split}
d_{z}(Z,P_{z})& =\frac{1}{4\left[1+\left(\tau_{2}+\tau_{3}\right)^{2}\right]}\left[-2\sqrt{2}k_{3}^{3}\tau_{3}^{2}\left(\tau_{2}+\tau_{3}\right)^{2}Z\right]+\sqrt{2}k_{3}^{3}\tau_{3}\left(\tau_{2}+\tau_{3}\right)Z\\
& -\frac{1}{\sqrt{2}}k_{3}^{3}\tau_{3}^{2}Z+\sqrt{2}k_{3}^{3}\tau_{3}\left(\tau_{2}+\tau_{3}\right)Z-\frac{1}{4\left[1+\left(\tau_{2}+\tau_{3}\right)^{2}\right]}\left[2\sqrt{2}k_{3}^{3}\tau_{3}^{2}Z\right]\\
& +P_{z}\left\{-4\sqrt{2}k_{3}^{3}\tau_{3} +\left(\tau_{2}+\tau_{3}\right)\left[2\sqrt{2}k_{3}^{3}\tau_{3}^{2}-4\sqrt{2}k_{3}^{3}\tau_{3}\left(\tau_{2}+\tau_{3}\right)\right]\right\}
\end{split}
\end{equation}

\begin{widetext}
\begin{equation}\label{dxa}
\begin{split}
d_{x}(X,P_{x}) &=\frac{1}{4\left[1+\left(\tau_{2}+\tau_{3}\right)^{2}\right]}\\
&\times\left(-2\sqrt{2}k_{2}^{3}\tau_{2}^{2}\left(\tau_{2}+\tau_{3}\right)^{2}X+2\sqrt{2}k_{2}^{3}\tau_{2}X\left\{4\left(\tau_{2}+\tau_{3}\right)-2\tau_{2}\left[1-\left(\tau_{2}+\tau_{3}\right)^{2}\right]\right\}\right)\\
& -\frac{1}{\sqrt{2}}k_{2}^{3}\tau_{2}^{2}X+P_{x}\left[2\sqrt{2}k_{2}^{3}\tau_{2}^{2}\left(\tau_{2}+\tau_{3}\right)-4\sqrt{2}k_{2}^{3}\tau_{2}\left(1+\tau_{2}^{2}+\tau_{2}\tau_{3}\right)\right]\\
& -\frac{1}{4\left[1+\left(\tau_{2}+\tau_{3}\right)^{2}\right]}\left(2\sqrt{2}k_{2}^{3}\tau_{2}^{2}X+8\sqrt{2}k_{2}^{3}\tau_{2}\tau_{3}X\right)
\end{split}
\end{equation}
\end{widetext}

\begin{widetext}
\begin{equation}\label{delza}
\begin{split}
\delta_{z}(Z,P_{z})& =-P_{z}2\sqrt{2}k_{3}^{3}\tau_{3}^{2}+2\sqrt{2}k_{3}^{3}\tau_{3}Z+2\sqrt{2}k_{3}^{3}\tau_{3}z_{0}
\end{split}
\end{equation}
\end{widetext}
\begin{widetext}
\begin{equation}\label{delxa}
\begin{split}
&\delta_{x}(X,P_{x}) =\frac{1}{4\left[1+\left(\tau_{2}+\tau_{3}\right)^{2}\right]}\left\{2\sqrt{2}k_{2}^{3}\tau_{2}X\left[2-2\left(\tau_{2}+\tau_{3}\right)^{2}+4\tau_{2}\left(\tau_{2}+\tau_{3}\right)\right]\right\}\\
& +\sqrt{2}k_{2}^{3}\tau_{2}X-P_{x}\left(2\sqrt{2}k_{2}^{3}\tau_{2}^{2}+4\sqrt{2}k_{2}^{3}\tau_{2}\tau_{3}\right)\\
& -\frac{1}{4\left[1+\left(\tau_{2}+\tau_{3}\right)^{2}\right]}\left\{-4\sqrt{2}k_{2}^{3}\tau_{2}X\left[2+2\tau_{2}\left(\tau_{2}+\tau_{3}\right)\right]\right\}+2\sqrt{2}k_{2}^{3}\tau_{2}x_{0}
\end{split}
\end{equation}
\end{widetext}

\bibliography{bibCSGE}

\end{document}